\documentclass[preprint,showpacs,preprintnumbers,amsmath,amssymb,endfloats]{revtex4}
\usepackage{graphicx}% Include figure files
\usepackage{dcolumn}% Align table columns on decimal point
\usepackage{bm}% bold math
\usepackage{amssymb}
\usepackage{amsmath}
\usepackage{longtable}
\usepackage{ulem}

\begin{document}

\title{Derivation of breakup probabilities from experimental     elastic
backscattering data}
\author{V.V.Sargsyan$^1$, G.G.Adamian$^{1}$, N.V.Antonenko$^1$, W. Scheid$^2$, and  H.Q.Zhang$^3$
}
\affiliation{$^{1}$Joint Institute for Nuclear Research, 141980 Dubna, Russia\\
$^{2}$Institut f\"ur Theoretische Physik der
Justus--Liebig--Universit\"at,
D--35392 Giessen, Germany\\
$^{3}$China Institute of Atomic Energy, Post Office Box 275, Beijing 102413,  China
}
\date{\today}

\pacs{25.70.Jj, 24.10.-i, 24.60.-k \\ Key words: sub-barrier capture, neutron transfer, quantum diffusion approach}

\begin{abstract}
  We suggest simple and useful
method  to extract breakup probabilities from the experimental
elastic   backscattering  probabilities in the reactions with toughly and weakly bound nuclei.
\end{abstract}

 \maketitle

%
%\section{Introduction}
%\section{Quasi-elastic and elastic backscattering - tools for search of  breakup process
%in reactions with weakly bound projectiles}

The fusion (capture) dynamics induced by loosely bound radioactive
ion beams  is  
 being extensively
studied \cite{Gomes,EPJSub4}.
However, the long-standing  question whether fusion (capture) is enhanced
or suppressed with these beams has not yet been answered unambiguously.
The study of the fusion reactions involving nuclei close to  the drip-lines has
led to contradictory results.
The lack of a clear systematic behavior \cite{EPJSub4,PRSGomes5} of the breakup probability
as a function of the target charge  requires   additional
experimental and theoretical studies.
The quasi-elastic backscattering
has been  suggested~\cite{EPJSub4}
as an alternative to investigate the breakup probability.
Since
the quasi-elastic experiments  are usually not as complex as the fusion (capture) and breakup
measurements, they are well suited to survey the breakup probability.

In the present report we will show that
by employing the experimental elastic backscattering data,
one can extract the breakup probabilities of  weakly bound nuclei.
So, new method  for the study of the breakup probability will be suggested.

There is a direct relationship between the capture,
  quasi-elastic scattering  and   breakup processes, since any losses from
the  elastic scattering and breakup channels contribute  directly to other channels 
(the conservation of the total reaction flux at given bombarding energy $E_{\rm c.m.}$ and angular momentum $J$):
\begin{eqnarray}
P_{qe}(E_{\rm c.m.},J)+P_{cap}(E_{\rm c.m.},J)+P_{BU}(E_{\rm c.m.},J)=\nonumber\\
=P_{el}(E_{\rm c.m.},J)+P_{rest}(E_{\rm c.m.},J)+P_{BU}(E_{\rm c.m.},J)=1,
\end{eqnarray}
where
$P_{rest}=P_{cap}+P_{in}+P_{tr}$,
$P_{qe}=P_{el}+P_{in}+P_{tr}$  
is the quasi-elastic scattering probability,
$P_{BU}$ is the breakup  probability, and
$P_{cap}$ is the capture  probability.
The quasi-elastic scattering
($P_{qe}$)  
is the sum of
all direct reactions, which include elastic ($P_{el}$),
inelastic ($P_{in}$), and
a few nucleon transfer ($P_{tr}$) processes.
In Eq. (1),  we neglect the deep-inelastic
collision process, since we are concerned with low energies.
 
Equation~(1) can be rewritten as
\begin{eqnarray}
\frac{P_{el}(E_{\rm c.m.},J)}{1-P_{BU}(E_{\rm c.m.},J)}+\frac{P_{rest}(E_{\rm c.m.},J)}{1-P_{BU}(E_{\rm c.m.},J)}
=P_{el}^{noBU}(E_{\rm c.m.},J)+P_{rest}^{noBU}(E_{\rm c.m.},J)=1,
\end{eqnarray} 
where
$$P_{el}^{noBU}(E_{\rm c.m.},J)=\frac{P_{el}(E_{\rm c.m.},J)}{1-P_{BU}(E_{\rm c.m.},J)}$$
and
$$P_{rest}^{noBU}(E_{\rm c.m.},J)=\frac{P_{rest}(E_{\rm c.m.},J)}{1-P_{BU}(E_{\rm c.m.},J)}$$
are the  elastic scattering and  other channels  probabilities, respectively,  in the absence of the
breakup process. From these expressions we obtain the useful formulas
\begin{eqnarray}
\frac{P_{el}(E_{\rm c.m.},J)}{P_{rest}(E_{\rm c.m.},J)}=\frac{P_{el}^{noBU}(E_{\rm c.m.},J)}{P_{rest}^{noBU}(E_{\rm c.m.},J)}
=\frac{P_{el}^{noBU}(E_{\rm c.m.},J)}{1-P_{el}^{noBU}(E_{\rm c.m.},J)}.
\end{eqnarray}
Using  Eqs.~(1) and (3), one can find the relationship between the breakup and  elastic scattering processes:
\begin{eqnarray}
P_{BU}(E_{\rm c.m.},J)=
%1-P_{el}(E_{\rm c.m.},J)[1+1/a]=
1-\frac{P_{el}(E_{\rm c.m.},J)}{P_{el}^{noBU}(E_{\rm c.m.},J)}.
\end{eqnarray}
 The last equation is the main result  of the present report.
Note that   similar  formula 
\begin{eqnarray}
P_{BU}(E_{\rm c.m.},J)=
%1-P_{qe}(E_{\rm c.m.},J)[1+1/a]=
1-\frac{P_{qe}(E_{\rm c.m.},J)}{P_{qe}^{noBU}(E_{\rm c.m.},J)} 
\end{eqnarray}
was derived in Ref. \cite{EPJSub4} to
relate  the breakup and quasi-elastic scattering processes.

The reflection  elastic or quasi-elastic backscattering probability
%\begin{eqnarray}
$P_{el,qe}(E_{\rm c.m.},J=0)=d\sigma_{el,qe}/d\sigma_{Ru}$
%\end{eqnarray}
for bombarding energy
$E_{\rm c.m.}$ and
angular momentum $J=0$ is given by the ratio of
the  elastic or quasi-elastic scattering differential cross section $\sigma_{el,qe}$  and
Rutherford differential cross section $\sigma_{Ru}$
at 180 degrees~\cite{Timmers}.
Employing Eq.~(4) or (5) and the  experimental  elastic or quasi-elastic backscattering data
in the reactions with toughly
and
weakly bound isotopes-projectiles and the same or almost the same compound nucleus,
one can extract the breakup probability of the exotic nucleus.
For example, using  Eq.~(4) or (5)  at backward angle,
the experimental 
$P_{el,qe}^{noBU}$[$^{4}$He+$^{A}$X] of the
$^{4}$He+$^{A}$X reaction with toughly bound nuclei (without breakup), and
the experimental  
$P_{el,qe}$[$^{6}$He+$^{A-2}$X]
of the $^{6}$He+$^{A-2}$X reaction with weakly bound projectile (with breakup), and assuming
approximate equality $V_b$($^{4}$He+$^{A}$X)$\approx V_b$($^{6}$He+$^{A-2}$X)
for the Coulomb barriers of   very asymmetric systems,
one can extract the breakup probability of the $^{6}$He:
\begin{eqnarray}
P_{BU}(E_{\rm c.m.},J=0)=
1-\frac{P_{el,qe}(E_{\rm c.m.},J=0)[^{6}{\rm He}+^{A-2}{\rm X}]}{P_{el,qe}^{noBU}(E_{\rm c.m.},J=0)[^{4}{\rm He}+^{A}{\rm X}]}
\end{eqnarray}
or 
\begin{eqnarray}
P_{BU}(E_{\rm c.m.},J=0)=
1-\frac{P_{el,qe}(E_{\rm c.m.},J=0)[^{6}{\rm He}+^{A-2}{\rm X}]}{P_{el,qe}^{noBU}(E_{\rm c.m.},J=0)[^{4}{\rm He}+^{A-2}{\rm X}]}.
\end{eqnarray}
Comparing the experimental elastic or quasi-elastic backscattering
probabilities in the presence and absence  of breakup
data in the reaction pairs
 $^{6}$He+$^{68}$Zn    and    $^{4}$He+$^{68,70}$Zn,
 $^{6}$He+$^{122}$Sn   and    $^{4}$He+$^{122,124}$Sn,
 $^{6}$He+$^{236}$U    and    $^{4}$He+$^{236,238}$U,
 $^{8}$He+$^{204}$Pb   and    $^{4}$He+$^{204,208}$Pb,
 $^{8}$Li+$^{207}$Pb   and    $^{7}$Li+$^{207,208}$Pb,
 $^{7}$Be+$^{207}$Pb   and   $^{10}$Be+$^{204,207}$Pb,
 $^{9}$Be+$^{208}$Pb   and   $^{10}$Be+$^{207,208}$Pb,
$^{11}$Be+$^{206}$Pb   and   $^{10}$Be+$^{206,207}$Pb,
 $^{8}$B+$^{208}$Pb    and   $^{10}$B+$^{206,208}$Pb,
 $^{8}$B+$^{207}$Pb    and   $^{11}$B+$^{204,207}$Pb,
 $^{9}$B+$^{208}$Pb    and   $^{11}$B+$^{206,208}$Pb,
$^{15}$C+$^{204}$Pb   and    $^{12}$C+$^{204,207}$Pb,
$^{15}$C+$^{206}$Pb   and    $^{13}$C+$^{206,208}$Pb,
$^{15}$C+$^{207}$Pb   and    $^{14}$C+$^{207,208}$Pb,
$^{17}$F+$^{208}$Pb   and    $^{19}$F+$^{206,208}$Pb,
leading to the same or almost the same corresponding compound nuclei,
one can analyse  the role of the breakup channels
in the reactions with the light weakly bound projectiles $^{6,8}$He, $^{8}$Li, $^{7,9,11}$Be,
$^{8,9}$B, $^{15}$C, and  $^{17}$F  at  energies near and above the Coulomb barrier.

One concludes that the   elastic or quasi-elastic backscattering technique
could be a very useful tool in the study of breakup.
The breakup probabilities can be  extracted    from the   elastic 
or quasi-elastic backscattering
probabilities of systems mentioned above.

We thank P.R.S.~Gomes and A.~L\'epina-Szily for fruitful discussions and suggestions.
This work was supported by   NSFC, RFBR, and JINR grants.
The IN2P3(France)-JINR(Dubna) and Polish - JINR(Dubna) Cooperation Programmes
are gratefully acknowledged.\newline
%
% BibTeX or Biber users please use (the style is already called in the class, ensure that the "woc.bst" style is in your local directory)
% \bibliography{name or your bibliography database}
%
% Non-BibTeX users please use
%

\end{document}